\documentclass[12pt]{article}

\usepackage{graphicx}
\usepackage{amsmath}
\usepackage{amssymb}
\usepackage{wrapfig}
\usepackage{indentfirst}
\usepackage{color}
\usepackage{subfigure}
\usepackage{hyperref}

\usepackage{cite}

\begin{document}

\vskip 0.5cm \centerline{\bf\Large Regge phenomenology and coherent photoproduction } 
\centerline{\bf\Large of $\mathbf{J/\psi}$  in peripheral heavy ion collisions}
\vskip 0.3cm
\centerline{L\'aszl\'o~Jenkovszky $^{a\star}$, Vladyslav~Libov$^{a\diamond}$, and Magno V. T. Machado$^{b\spadesuit}$}

\vskip 1cm

\centerline{$^a$ \sl Bogolyubov ITP,
National Academy of Sciences of Ukraine, Kiev
03143 Ukraine}
\centerline{$^b$ \sl HEP Phenomenology Group, CEP 91501-970, Porto Alegre, RS, Brazil}
\vskip
0.1cm

\begin{abstract}\noindent
By using models based on Regge phenomenology we analyse the coherent photoproduction of charmonium in peripheral heavy-ion collisions at the Large Hadron Collider (LHC).
The centrality dependence is investigated and compared to the experimental results for coherent $J/\psi$ production in lead-lead LHC runs at the energies of 2.76 and 5.02~TeV.
Theoretical uncertainties and possible limitations of the formalism are also discussed.
\end{abstract}

\vskip 0.1cm

$
\begin{array}{ll}
^{\star}\mbox{{\it e-mail address:}} &
   \mbox{jenk@bitp.kiev.ua} \\
^{\diamond}\mbox{{\it e-mail address:}} &
   \mbox{vladyslav.libov@gmail.com} \\
 ^{\spadesuit}\mbox{{\it e-mail address:}} &
\mbox{magnus@if.ufrgs.br}\\  

\end{array}
$


\section{Introduction}\label{Int}

High precision data on exclusive quarkonium production in semi-central and peripheral  heavy ion  collisions became recently available \cite{ALICE:2015mzu,Massacrier:2019lmf,LHCb:2021hoq,Zha:2017etq}. Systematically, event excess over the expected hadronic $J/\psi$ production at very low transverse momentum, $p_T$,  both at Large Hadron Collider (LHC) and Relativistic Heavy Ion Collider (RHIC) energies has been interpreted as due to coherent quarkonium photoproduction in hard collisions with nuclear overlap. At RHIC,  an excess in the $J/\psi$ yield at low-$p_T$ has been observed by the STAR Collaboration in AuAu collsions at $\sqrt{s_{\mathrm{NN}}}=200$ GeV and UU collisions at $\sqrt{s_{\mathrm{NN}}}=193$ GeV   at mid-rapidity \cite{Zha:2017etq}. ALICE collaboration has also measured an excess at very low-$p_T$ in the
yield of $J/\psi$ in PbPb collisions at $\sqrt{s_{\mathrm{NN}}}=2.76$ TeV \cite{ALICE:2015mzu}. The excess observed in peripheral heavy ion collisions correspond to the centrality classes of $30-50\%$, $50-70\%$ and $70-90\%$, respectively.  Recently, ALICE  reported similar  results for PbPb collisions at $\sqrt{s_{\mathrm{NN}}}=5.02$ TeV. Measurements  are presented at mid and forward rapidity and for the same classes of centrality. Moreover, the photoproduction at low transverse momentum has been studied in peripheral PbPb collisions by LHCb collaboration at 5 TeV \cite{LHCb:2021hoq}. The yield of coherently photoproduced meson has been extracted as a function of rapidity and transverse momentum  in intervals of the number of participant nucleons in  rapidity range $2.0<y<4.5$. 

From the theoretical point of view the coherent photoproduction of quarkonium in peripheral heavy ion collisions has been addressed in a number of works \cite{Contreras:2016pkc,Klusek-Gawenda:2015hja,GayDucati:2017ksh,GayDucati:2018who,Zha:2017jch,Zha:2018ytv,Zha:2018jin,Shi:2017qep,Yu:2021sxg}. In Ref. \cite{Contreras:2016pkc}  the coherent photonuclear cross section has been extracted from coherent $J/\psi$ production using the ALICE measurements from peripheral and ultra-peripheral PbPb collisions at 2.76 TeV. The equivalent photon approximation (EPA) has been assumed and the effective photon flux for a given centrality class was taken from Ref. \cite{Klein:1999qj}. A different approach was considered in Ref. \cite{Klusek-Gawenda:2015hja}, where absorption effects were introduced by modifying the photon flux in the impact parameter space. Still using EPA approximation, this is achieved by imposing many geometrical conditions on impact parameters between $\gamma$ and Pb and between colliding lead nuclei. The photonuclear cross section, $\sigma (\gamma +A \rightarrow V+A)$, is described through vector dominance model and nuclear effects given by Glauber approach. Similar study was considered in Refs. \cite{GayDucati:2017ksh,GayDucati:2018who}, where on the other hand the photonuclear cross section is computed in the QCD color dipole framework. The investigation done in Refs. \cite{Zha:2017jch,Zha:2018ytv,Zha:2018jin} addresses the role played by the participation of spectator and non-spectator
nucleons  in the coherent scattering. It was found that the strong interactions in the overlapping region of incoming nuclei modify the  $J/\psi$ coherent production. Different coupling scenarios
for photons and pomerons with the nucleus in presence
of hadronic interactions were considered. It was pointed out that the  destructive interference between photoproduction on ions moving in opposite directions plays an important role. Interestingly,   modifications of an anisotropic Quark Gluon Plasma (QGP) on photoproduced mesons have been studied in Ref. \cite{Shi:2017qep}. The  spatial configurations of produced quarkonia at transverse coordinate $\vec{r}_{\perp}$ is given by their distribution  over the nucleus surface with a normalized distribution, $f^{\mathrm{
norm}}(\vec{r}_{\perp} )$, which is  proportional to the square
of target thickness function. This is compared with the quarkonia uniformly distributed over the target nuclear surface in order to single out the theoretical uncertainties. Study presented in Ref.  \cite{Yu:2021sxg} considers standard EPA approximation and color dipole framework. Parton saturation models for the dipole-proton cross section were utilized and the numerical calculation compared to the RHIC data.

In this study coherent $J/\psi$ photoproduction in peripheral collisions at the LHC will be scrutinized. In our calculations we consider the Vector Dominance Model (VDM) and Glauber multiple scattering formalism which is quite appealing due to its simplicity and direct connection with Regge phenomenology and Reggeon Field Theory (RFT) \cite{Jenkovszky:2018itd}. The theoretical input is given by the single-component Reggeometric Pomeron model \cite{FFJS1,FFJS2} and the Soft Dipole Pomeron model  \cite{Fiore:1998jx,Martynov:2002ez}.  The present study extends our earlier work~\cite{Fiore:2014oha,Fiore:2014lxa,Fiore:2015yya,Jenkovszky:2021sis} where we have addressed the rapidity distribution for $J/\psi$ and $\psi (2S)$ mesons in proton-proton collisions at the LHC for the center-of-mass energies of 7, 8 and 13 TeV.  In Ref. \cite{Jenkovszky:2021sis} the coherent $J/\psi$ and $\psi (2S)$ photoproduction  in ultraperipheral PbPb (at  $\sqrt{s_{\mathrm{NN}}}=2.76$ TeV and  $\sqrt{s_{\mathrm{NN}}}=5.02$ TeV) and XeXe collisions (at  $\sqrt{s_{\mathrm{NN}}}=5.44$ TeV and  $\sqrt{s_{\mathrm{NN}}}=5.86$ TeV)  have been analyzed. Rapidity and  transverse momentum distributions were predicted and numerical results were generally consistent with experimental data. This paper is organized as follows. In Sec. \ref{sec:theory} we shortly review the exclusive vector meson production in the context of the Reggeometric and Soft Dipole Pomeron models in the process $\gamma+p\rightarrow  V+p$. In Subsection \ref{sec:photonuclear} the formalism to obtain the photonuclear cross section for the process $\gamma+A\rightarrow V+A$ is presented. The numerical results for peripheral PbPb reactions at different centrality classes are compared to experimental measurements in Sec. \ref{sec:results}, where a discussion on the theoretical uncertainties is done. In the last section we summarize the main results.

\section{Theoretical formalism}
\label{sec:theory}
Let us start by introducing the models for the quarkonium photoproduction describing the exclusive process, $\gamma +p\rightarrow V+p$. Two models based on Regge phenomenology are considered. The first one is the Reggeometric Pomeron model and the  second is the Soft Dipole Pomeron model. An advantage is that they are able to describe also the electroproduction data where the virtuality dependence is taken into account by $\widetilde Q^2$-dependent factors.  Accordingly, the measure of the hardness, $\widetilde Q^2=Q^2+M_V^2$, is the sum of the squared photon virtuality $Q^2$ and the squared mass $M^2_V$ of the produced vector meson.   

\begin{figure}[t]
\centering
\includegraphics[width=.8\textwidth]{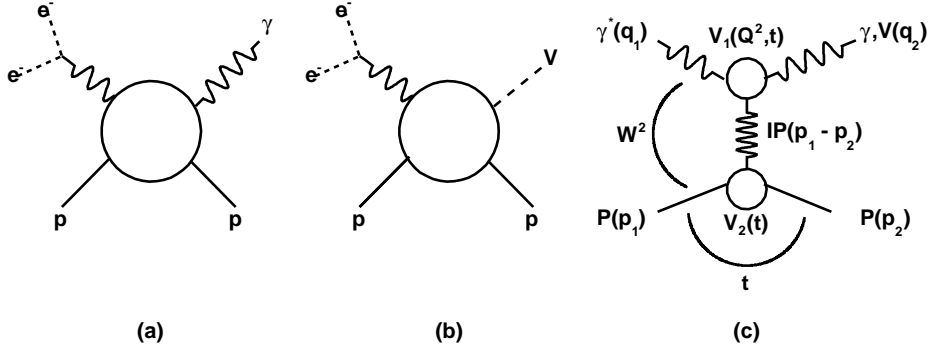}
\caption{Diagrams of Deeply Virtual Compton scattering (DVCS) (a) and vector meson production (VPM) (b) in $e^{\pm}p$ scattering; (c) DVCS (VMP) amplitude in a Regge-factorized form including representation for particle vertices, $V_{1,2}$.}
\label{fig:diagrams}
\end{figure}

The scattering amplitude associated to the single-component Reggeometric Pomeron (RP)  in a given scale  $\widetilde Q^2$ is given by \cite{FFJS1,FFJS2}:
\begin{eqnarray}
{\cal{A}}_{\mathrm{RP}}(Q^2,s,t)=\frac{\widetilde{A_0}}{\left(1+\frac{\widetilde{Q^2}}{{Q_0^2}}\right)^{n}} e^{-\frac{i\pi\alpha(t)}{2}}\left(\frac{s}{s_0}\right)^{\alpha(t)} e^{(B_0/2)t}, \quad B_0(\widetilde{Q^2})= 4\left(\frac{a}{\widetilde{Q^2}}+\frac{b}{2m_N^2}\right),
\label{amplitudesingle}
\end{eqnarray}
where the exponent in the exponential factor in Eq. (\ref{amplitudesingle}) reflects its geometrical nature and where  $\alpha (t)$ is the Pomeron trajectory. The center-of-mass energy of photon-nucleon system is denoted as $\sqrt{s}=W_{\gamma p}$ and the  quantities $a/\widetilde Q^2$ and $b/2m_N^2$ in  (\ref{amplitudesingle})  correspond to the effective sizes of upper and lower vertices in Fig. \ref{fig:diagrams}-c. The elastic differential cross section takes the form:
\begin{eqnarray}
\frac{d\sigma_{el}}{dt}=\frac{A_0^2}{\left(1+\frac{\widetilde{Q^2}}{{Q_0^2}}\right)^{2n}}\left(\frac{s}{s_0}\right)^{2(\alpha(t)-1)}\exp \left[B_0(\widetilde{Q^2})\,t\right].
\end{eqnarray}
 
The Reggeometric Pomeron model naturally applies to photoproduction ($Q^2=0$) limit, where $\tilde{Q}^2=M_V^2$. The  photoproduction cross section is given by  \cite{FFJS1,FFJS2}:
\begin{eqnarray}
\sigma_{\gamma p \to V p} (W_{\gamma p},\widetilde Q^2) & = &\frac{A_0^2}{\left(1+\widetilde Q^2/\widetilde Q^2_0\right)^{2n}}\frac{\left(W_{\gamma p}/W_0\right)^{4(\alpha_0-1)}}{B_V\left(W,\widetilde Q^2 \right)}, \label{eq:jpsiproduction}\\
 B_V \left(W_{\gamma p},\widetilde Q^2 \right) & = &  4\left[\alpha'\ln(W_{\gamma p}/W_0)+\left(\frac{a}{\widetilde Q^2}+\frac{b}{2m_N^2}\right)\right].
\label{eq:bvslope}
\end{eqnarray}
The parameters, fitted \cite{FFJS1}  to the $J/\psi$ photoproduction data, are presented in Table \ref{tab:1}. The analysis done in \cite{FFJS1} also includes the production of $\phi$ and Deeply Virtual Compton scattering (DVCS). 

\begin{table}[t]
  \centering
   \footnotesize
   \caption{Values of the parameters for the Reggeometric Pomeron model  \cite{FFJS1}  fitted to data on $J/\psi$ production at HERA. }
   \label{tab:1}
  \begin{tabular}{|c|c|c|c|c|c|c|}
     \hline
                 $A_0$ $\left[\frac{\sqrt\text{nb}}{\text{GeV}}\right]$
                 &$\widetilde{Q^2_0}$ $\left[\text{GeV}^2\right]$&   $n$
                 &$\alpha_{0}$& $\alpha'$  $\left[\text{GeV}^{-2}\right]$
                 &$a$&$b$ \\ \hline 
         29.8 $\pm$ 2.8 & 2.1 $\pm$ 0.4      &1.37 $\pm$ 0.14 &1.20 $\pm$ 0.02 & 0.17 $\pm$ 0.05& 1.01 $\pm$ 0.11 & 0.44 $\pm$ 0.08  \\ 
 \hline
        \end{tabular}
 \end{table}
 
Let us now move to the Soft Dipole Pomeron model (SDP) applied to  diffractive
photoproduction of quarkonium \cite{Fiore:1998jx,Martynov:2002ez}.  The corresponding invariant scattering amplitude is written  as follows~\cite{Fiore:1998jx}:
\begin{eqnarray}
{\cal{A}}_{\mathrm{SDP}}(s,t) = i\left(\frac{-is}{s_0}\right)^{\alpha(t)-1}\left\{G_1(t)+G_2(t)\left[\ln \left(\frac{s}{s_0}\right)-i\frac{\pi}{2}\right]\right\},
\end{eqnarray}
where the residua of the simple, $G_1$, and double pole, $G_2$, are given by,
\begin{eqnarray}
G_1(t)=A_1e^{bt}(1+h_1t), \quad G_2(t)=A_2e^{bt}(1+h_2t)-\gamma .
\label{residua}
\end{eqnarray}
Concerning the quantities appearing in Eq. (\ref{residua}) one has the 
standard $p I\!\!P p$ vertex $\sim e^{b t}$, 
with $b=2.25~GeV^{-2}$ determined from the $p p$ scattering, 
and the $\gamma I\!\!P V$ vertex which is parameterized by a simple polynomial  with a free parameter $h_1$. For simplicity, in Ref. \cite{Fiore:1998jx} a linear trajectory for the 
Pomeron, $\alpha(t)=1 + 0.25 t$, was considered. 

The elastic differential cross section reads~\cite{Fiore:1998jx} as
\begin{eqnarray}
\frac{d\sigma_{el}}{dt}&=&\left(\frac{s}{s_0}\right)^{2\alpha(t)-2}\left\{\left[G_1(t)+G_2(t)\ln{\left(\frac{s}{s_0}\right)}\right]^2+
\frac{\pi^2}{4} G_2^2(t)\right], \\
\left. \frac{d\sigma_{el}}{dt}\right|_{t=0} &= &
\left[A_1+(A_2-\gamma)\ln{\left(\frac{s}{s_0}\right)}\right]^2+\frac{\pi^2}{4} (A_2-\gamma)^2.
\end{eqnarray}
The parameters $A_1, \,A_2, \,h_1,\, h_2$ and $\gamma$ were fitted to experimental measurements at HERA~\cite{Fiore:1998jx}. The values of the parameters for $J/\psi$ are presented in Table \ref{tab:2}. 

\begin{table}[t]
  \centering
   \footnotesize
   \caption{Values of the parameters for the Soft Dipole Pomeron model  \cite{Fiore:1998jx}  fitted to data on $J/\psi$ production at HERA. }
   \label{tab:2}
  \begin{tabular}{|c|c|c|c|c|c|}
     \hline
                 $A_1$ $\left[\frac{\sqrt{\mu\text{b}}}{\text{GeV}}\right]$
                 &$A_2$ $\left[\frac{\sqrt{\mu\text{b}}}{\text{GeV}}\right]$&   $h_1$ [GeV$^{-2}$]
                 &$h_2$  [GeV$^{-2}$]& $\gamma$ $\left[\text{GeV}^{-2}\right]$
                 &$s_0$ $\left[\text{GeV}\right]$ \\ \hline 
         0.27523 & 0.091278     & -0.80606  & 0 & 0 & 30 \\ 
 \hline
        \end{tabular}
 \end{table}

In next subsection we investigate the coherent nuclear scattering using as input the two models discussed above. In particular, we calculate the cross section $\sigma (\gamma +A\rightarrow V+A)$ as a function of the centre of mass energy for the photon-nucleus system, $W_{\gamma A}$. The nuclear effects are based on the Glauber formalism.

\subsection{Coherent charmonium production in heavy ion collisions}
\label{sec:photonuclear}

In what follows we investigate the coherent quarkonium  production in nuclear targets ($J/\psi$ in particular). Here the nuclear effects for the process, $\gamma +A\rightarrow V+A$ (coherent scattering),  are described by vector dominance model (VDM) and Glauber approach for multiple scattering. Using the classical mechanics Glauber formula for multiple scattering of vector meson in the
nuclear medium \cite{Bauer:1977iq} the differential cross section is given by:
\begin{eqnarray}
\left. \frac{\mathrm{d}\sigma\left( \gamma + A \to J/\psi+ A \right)}{\mathrm{d}t}\right|_{t=0}
 =   \frac{\alpha_{em}}{4 f^2_{J/\psi}}\left\{\int \mathrm{d}^2 \textbf{b} 
\left[ 1-\exp\left( -\sigma_{tot}\left( J/\psi p \right) T_A\left(\textbf{b} \right) \right) \right]\right\}^2,
\label{glauberVMD}
\end{eqnarray}
where $T_A(b)$ is the nuclear thickness function and $f_{J/\Psi}$ is the vector-meson coupling. The value  $f^2_{J/\Psi}/4\pi = 10.4$ will be considered. The approach above can been extended in the generalized vector dominance model (GVDM) \cite{Frankfurt:2003wv}. The input for the Glauber model calculation in Eq. (\ref{glauberVMD}) is the cross section for the process $J/\psi+p\rightarrow J/\psi+p $, which is given by:
\begin{eqnarray}
\sigma_{tot}\left( J/\psi p \right)= \sqrt{\frac{4f^2_{J/\psi}}{ \alpha_{em}} \left.
\frac{\mathrm{d}\sigma\left( \gamma + p \to J/\psi + p
  \right)}{\mathrm{d}t}\right|_{t=0}},
  \label{sigpsip}
\end{eqnarray}
where the predictions from the Reggeometric Pomeron  and Soft Dipole Pomeron models will be introduced in Eq.  (\ref{sigpsip}) .
The integrated cross section  can be written as:
\begin{eqnarray}
\sigma (\gamma + A \rightarrow J/\psi + A) = 
\left. \frac{d\sigma (\gamma + A \rightarrow  J/\psi + A)}{dt}\right|_{t=0}
\int\limits_{t_{min}}^{\infty} \mathrm{d}|t| \, \left| F_A\left(t\right) \right|^2, 
\label{eq:sigtotgammaA}
\end{eqnarray}
where $F_A$ is the nuclear form factor. In our calculation we consider an analytic form factor given by a hard sphere of radius, $R_A =1.2A^{1/3}$ fm, convoluted with a Yukawa potential with range $a$ \cite{Davies:1976zzb}, which is sufficient for practical use \cite{Klein:1999qj}:
\begin{eqnarray}
 F_A(|k|)  =  \frac{4\pi\rho_0}{A |k^3|} \left( \frac{1}{1+a^2k^2} \right) \left[ \sin{(|k| R_A)} - |k| R_A\cos{(|k| R_A)}  \right], 
\end{eqnarray}
where $A$ is the mass number, $k$ is the momentum transfer, $\rho_0 = A/\frac{4}{3}\pi R_A^3$ and $a = 0.7$ fm.

\begin{figure}[t]
\centering
 \includegraphics[width=.6\textwidth]{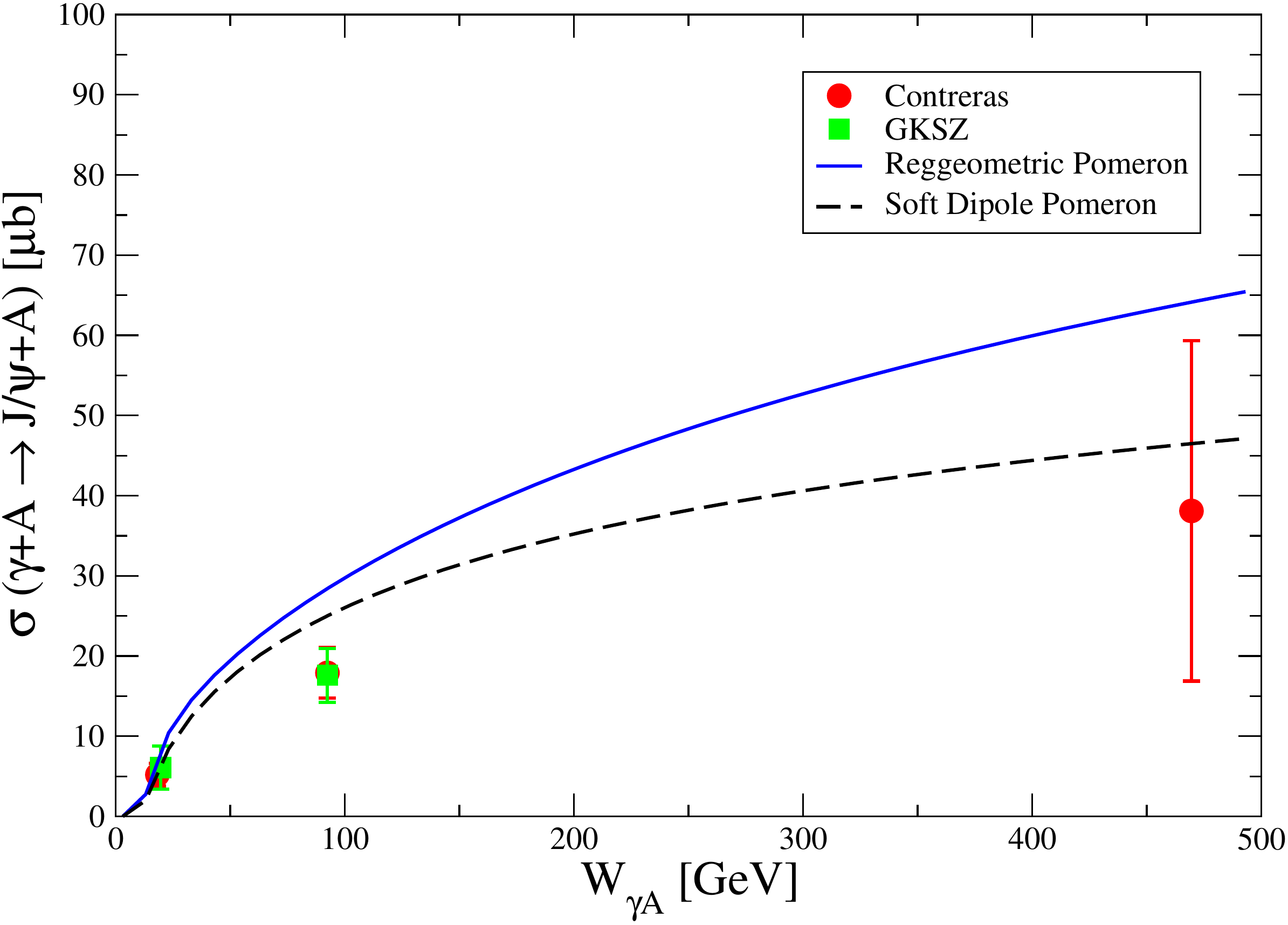}
 \caption{The coherent nuclear cross section, $\sigma (\gamma + A\rightarrow J/\psi + A)$,  as a function of the corresponding photon-nucleus energy. Theoretical predictions for Reggeometric Pomeron (solid  line) and Soft Dipole Pomeron (long-dashed line) models are shown. }
 \label{fig:photonuclearxs}
\end{figure}

In order to assess the robustness of the Pomeron exchange models, we compare our predictions to the coherent nuclear scattering cross sections, $\sigma (\gamma +A \rightarrow J/\psi + A)$, extracted from the ultra-peripheral nucleus-nucleus collisions at RHIC and LHC. Figure~\ref{fig:photonuclearxs}
shows the photonuclear cross section for $J/\psi$ production as a function of photon-nucleus energy, $W_{\gamma A}$. The extracted cross sections are from  Ref. \cite{Contreras:2016pkc} (labeled
Contreras) and Ref. \cite{Guzey:2013xba} (labeled GKSZ) with  the data description being quite reasonable. The Soft Dipole Pomeron model describes better the normalization and energy behavior. It is expected that models based on Glauber approach underestimate the nuclear suppression as the nuclear shadowing is not so intense in those cases (see discussion in Refs. \cite{Guzey:2013xba,Guzey:2020ntc}).

\section{Peripheral collisions - results and discussions }
\label{sec:results}

\begin{figure}[t]
\centering
 \includegraphics[width=.6\textwidth]{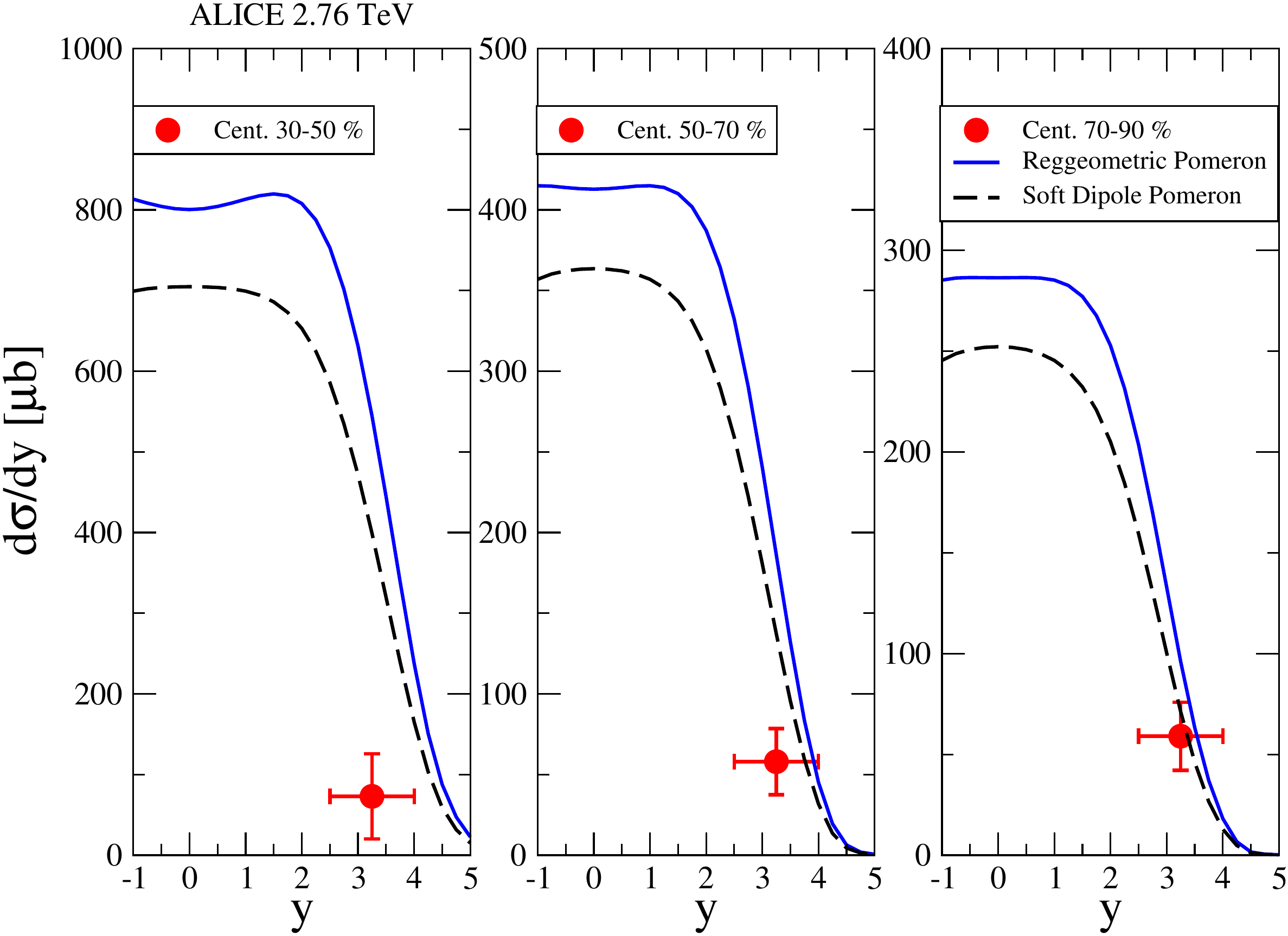}
 \caption{Rapidity distribution of coherently photoproduced $J/\psi$ at forward rapidities in Pb-Pb collisions at $\sqrt{s_{NN}}= 2.76$ TeV. ALICE data \cite{ALICE:2015mzu} are presented in the centrality range 30-50\% (left), 50-70\% (middle) and 70-90\% (right), compared to the theoretical calculations. Reggeometric Pomeron predictions are represented by solid curves and those for the Soft Dipole Pomeron by the long-dashed curves. }
 \label{fig:ALICE276centrality}
\end{figure}

Now we employ the equivalent photon approximation (EPA) where the transverse electromagnetic fields are treated as a bunch of quasi-real photons moving in the longitudinal direction. By integrating the photonuclear cross section over the photon
spectrum the total cross section is found to be:
\begin{eqnarray}
\sigma(A+A\rightarrow A+V+A) = \int d\omega \frac{dN_{\gamma}}{d\omega}
\,\sigma(\gamma + A\rightarrow V+A),
\end{eqnarray}
where $\omega$ is the photon energy. The total photon flux modulated by the non-interaction probability is given by \cite{Contreras:2016pkc,Klein:1999qj}:  
\begin{eqnarray}
\frac{dN_{\gamma}}{d\omega} =  
\int_{b_{min}}^{b_{max}} d^2\vec{b}\, \left(1-P_{NH}(b)\right)
    \int_0^{R_A}  \frac{rdr}{\pi R_A^2} 
\int_0^{2\pi} d\phi \frac{d^3N_{\gamma}(k,b+r\cos(\phi))}{d\omega d^2r},
\label{photonfluxcent}
\end{eqnarray}
where the  $b$ integral runs over impact parameter range $(b_{min},b_{max})$ according to the centrality determination (see below for more details).  

The interaction probability at a given impact parameter is
related to the overlap function $T_{AB}(b) = \int d^2\vec{r}\, T_A(\vec{r}) T_B(\vec{r}-\vec{b})$.  In our calculation the nuclear thickness function $T_A(\vec{r})$ is obtained using  nuclear
density given by the Woods-Saxon distribution. The experimental condition for peripheral collisions is that the nuclei undergo a hadronic collision. Accordingly, the probability of having no hadronic interactions is given by,
\begin{eqnarray}
P_{NH}(b)=\exp \left[-T_{AA}(b)\sigma_{NN}\right],
\label{nointeractions}
\end{eqnarray}
where the values $\sigma_{NN}= 61.8 $ mb at 2.76 TeV and $\sigma_{NN}=67.6 $ mb at 5.02 TeV have been considered \cite{Loizides:2017ack}.

Before proceeding, some words on the technical details are in order. In the comparison with ALICE data, the range on $b$ for a given centrality,  $(b_{min},b_{max})$, is taken from Table 1 of Ref. \cite{ALICE:2013hur} for $\sqrt{s_{NN}}=2.76$ TeV and Refs. \cite{ALICE:2018tvk,ALICE:2015juo} and Table 1 of Ref. \cite{ALICE-PUBLIC-2015-008} for $\sqrt{s_{NN}}=5.02$ TeV. In the comparison with LHCb data, we follow the centrality determination presented in Ref. \cite{LHCb:2021ysy} and further details given at the  LHCb public pages \cite{LHCB-PAPER-2020-043-webpage}.  For practical purposes, by using a Woods-Saxon distribution  the impact parameter range can be directly extracted from Ref. \cite{Loizides:2017ack} referred before.

The photon flux from a relativistic heavy nucleus is given by the
Weizs\"acker-Williams approach.  The photon flux at
a distance $r$ from the nucleus and for a target frame
$\gamma$-factor is \cite{Klein:1999qj},
\begin{eqnarray}
\frac{d^3N_{\gamma}(\omega, r)}{d\omega d^2r}= \frac{Z^2\alpha_{em} u^2}{\pi^2\omega r^2} \left[  K_1^2(u) + \frac{1}{\gamma^2}  K_0^2(u)\right]
\label{drgammaflux}
\end{eqnarray}
where $Z$ is the nuclear charge and  $K_{0,1}$ are the 
modified Bessel function of second kind with $u=\omega r/\gamma$.

\begin{figure}[t]
\centering
 \includegraphics[width=.6\textwidth]{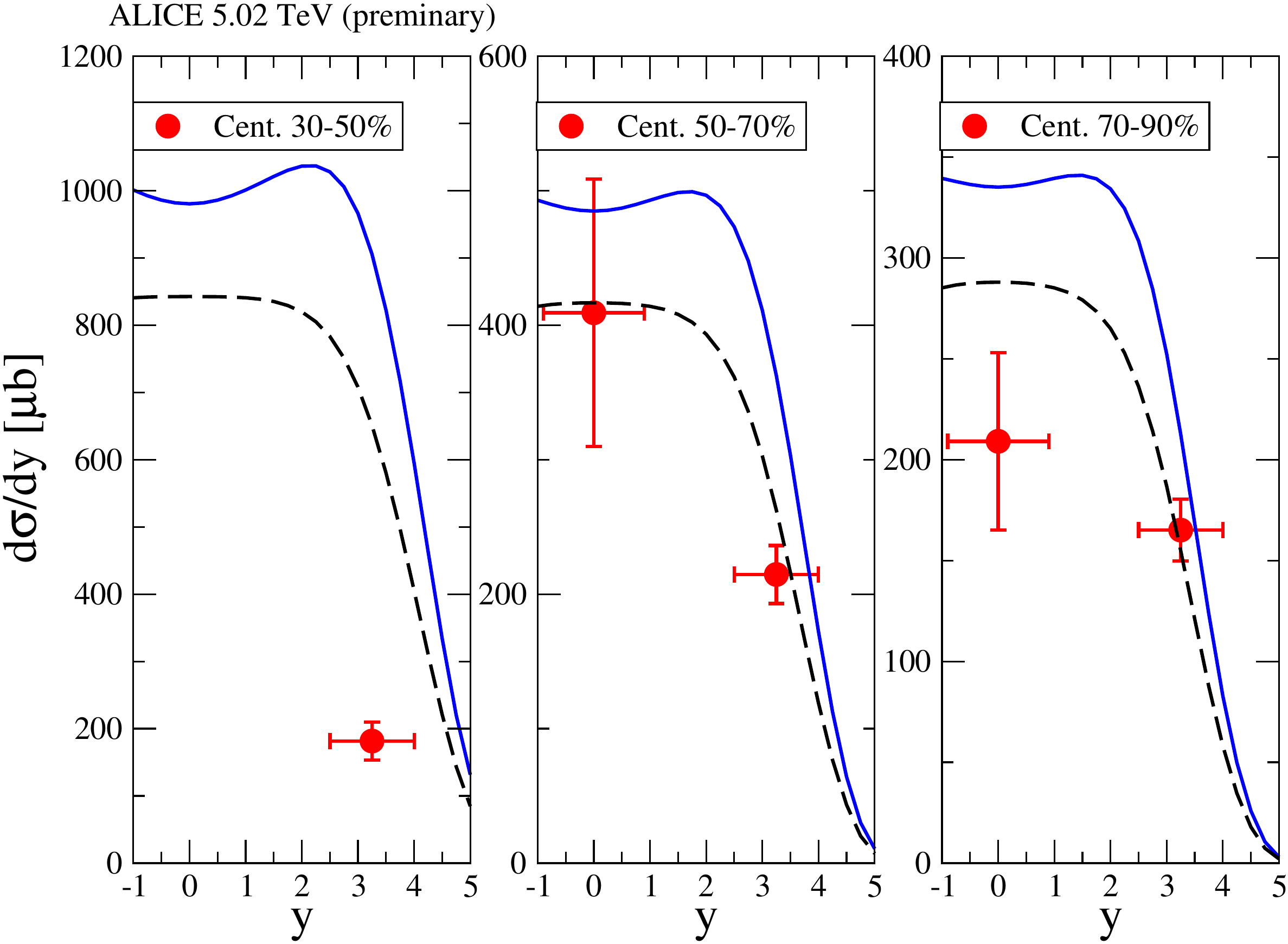}
 \caption{Rapidity distribution of coherently photoproduced $J/\psi$ at forward rapidities in Pb-Pb collisions at $\sqrt{s_{NN}}= 5.02$ TeV. Preliminary ALICE data \cite{Massacrier:2019lmf} are presented in the centrality range 30-50\% (left), 50-70\% (middle) and 70-90\% (right), compared to theoretical calculations. Same notation as Fig. \ref{fig:ALICE276centrality}.}
  \label{fig:ALICE502centrality}
\end{figure}

In this section the numerical results are presented and compared to recent experimental measurements. Let us start with the rapidity distribution of $J/\psi$ production in lead-lead (PbPb) collision at the LHC energies. The coherent production of $J/\psi$ in $AA$ collisions is straightforwardly computed in the EPA approximation. The rapidity distribution is given by:
\begin{eqnarray}
\frac{d\sigma (A+A\rightarrow A+V+A)}{dy}=\omega_+\frac{dN_{\gamma}(\omega_+)}{d\omega}\sigma_{\gamma A\rightarrow VA}(\omega_+) +\omega_-\frac{dN_{\gamma}(\omega_-)}{d\omega}\sigma_{\gamma A\rightarrow VA}(\omega_-),
\end{eqnarray}
where the rapidity $y$  of the vector meson $V$ is related to the center-of-mass energy of the photon-nucleus system, $W_{\gamma A}^2 =\sqrt{s_{\mathrm{NN}}}M_Ve^{y}$. The quantity $\sqrt{s_{\mathrm{NN}}}$ is the center-of-mass energy per nucleon pair in the AA system and $\omega_{\pm} = \frac{M_V}{2}e^{\pm y}$. The photon flux for a given centrality interval will be described by expression in Eq. (\ref{photonfluxcent}) and the photonuclear cross section is computed using Eq. (\ref{eq:sigtotgammaA}).

\begin{figure}[t]
\centering
 \includegraphics[width=.6\textwidth]{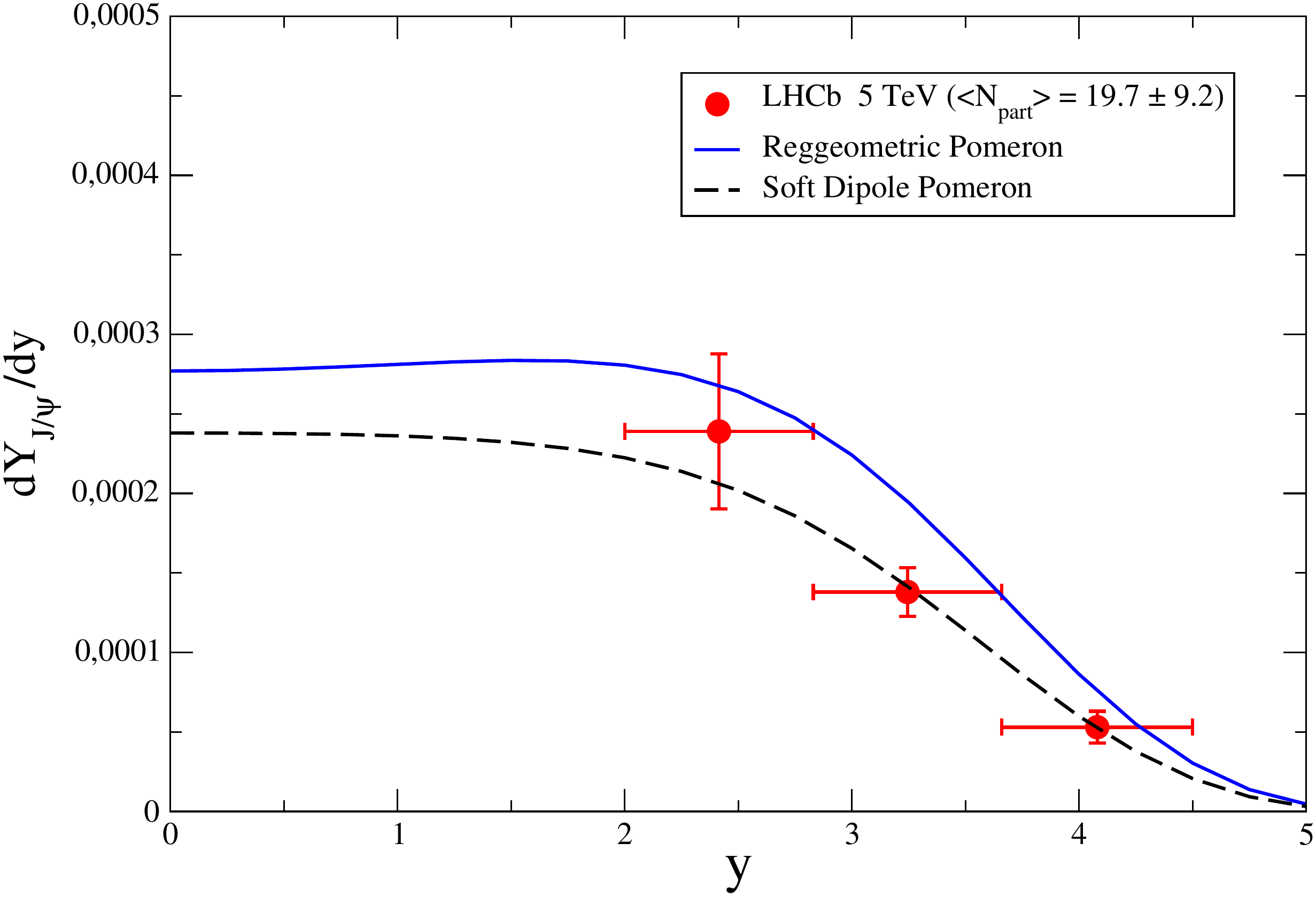}
 \caption{The differential $J/\psi$ photoproduction yields as a function of the rapidity in peripheral PbPb collisions for $\langle N_{part} \rangle =19.7 \pm 9.2$. Data from  LHCb collaboration at 5 TeV \cite{LHCb:2021hoq}. Same notation as previous figures.}
  \label{fig:LHCB502centrality}
\end{figure}

In Fig. \ref{fig:ALICE276centrality} results are shown for coherent nuclear $J/\psi$ production for PbPb peripheral collisions at $\sqrt{s_{\mathrm{NN}}} = 2.76$ TeV ($p_T<0.3$ GeV). The rapidity distribution of the meson is depicted as a function of the centrality bin. The experimental measurements from ALICE collaboration \cite{ALICE:2015mzu} are presented  in the centrality intervals of 30-50\%, 50-70\% and 70-90\%. The predictions for the Reggeometric Pomeron (RP) are labeled by the solid curves whereas the ones for Soft Dipole Pomeron are labeled by the long-dashed curves, respectively. Both models produce generally similar results, with RP systematically overshooting the SDP. The predictions overestimate the experimental measurements at semi-central  centralities and are in agreement with data for the more peripheral ones. The standard photon flux we used systematically overestimates the cross section in semi-central collisions. The reason is related to the absorption effects in the region of overlapping colliding nuclei in b-space (small-$b$ for semi-central centralities). The size of these absorption corrections is model-dependent and in general they are implemented as a modification of the photon flux. We enumerate some of the phenomenological approaches in literature in what follows. Actually, the corrections in the flux suppress the differential cross section $d\sigma/db (A+A\rightarrow A+V+A)$ for $b<<2R_A$.  For instance, we quote the Figs. 8 and 9 of Ref. \cite{Klusek-Gawenda:2015hja}  as an illustration of the referred effects. On the other hand, in case of peripheral collisions the formalism presented here  produces numerically similar results as other approaches to compute the effective photon flux as discussed below.

In Ref. \cite{Klusek-Gawenda:2015hja} the photoproduction mechanism for peripheral and semi-central collisions has been proposed.  It is  assumes that the whole nucleus produces photons and they must interact  with the other nucleus in order to produce a heavy meson. The treatment of  the region of overlapping colliding nuclei in the $b$-space is based on absorption effects which are taken into account by changing the effective photon fluxes.  This is enforced by imposing  geometrical conditions on impact
parameters between $\gamma$ and $A$ and/or between scattering nuclei.
The resulting modification describes different centrality bins adequately. Other works follow the same procedure, however in Refs. \cite{GayDucati:2017ksh,GayDucati:2018who} an effective photonuclear cross section is proposed (comparison between the different modifications in the photon flux and photonuclear cross section is performed). Another approach has been presented in Refs. \cite{Zha:2017jch,Zha:2018ytv,Zha:2018jin} where two scenarios are assumed  in which the coherence of the vector meson production is influenced or not by scattering with the overlap region of the colliding nuclei.

 In Fig. \ref{fig:ALICE502centrality} results are shown for the energy of  $\sqrt{s_{\mathrm{NN}}} = 5.02$ TeV. The preliminary  measurements from ALICE  collaboration \cite{Massacrier:2019lmf} are presented with the cut $p_T<0.3$ GeV. The notation is the same as previous figure. For the centralities bins of 50-70\% and 70-90\% both models describe data within the errors (summed into quadrature), with SDP being somewhat preferred. The predictions follow similar trend as observed at 2.76 TeV.  In general, models are suitable to predict the magnitude of the forward rapidity measurements in peripheral collisions.
 
\begin{figure}[t]
\centering
 \includegraphics[width=.6\textwidth]{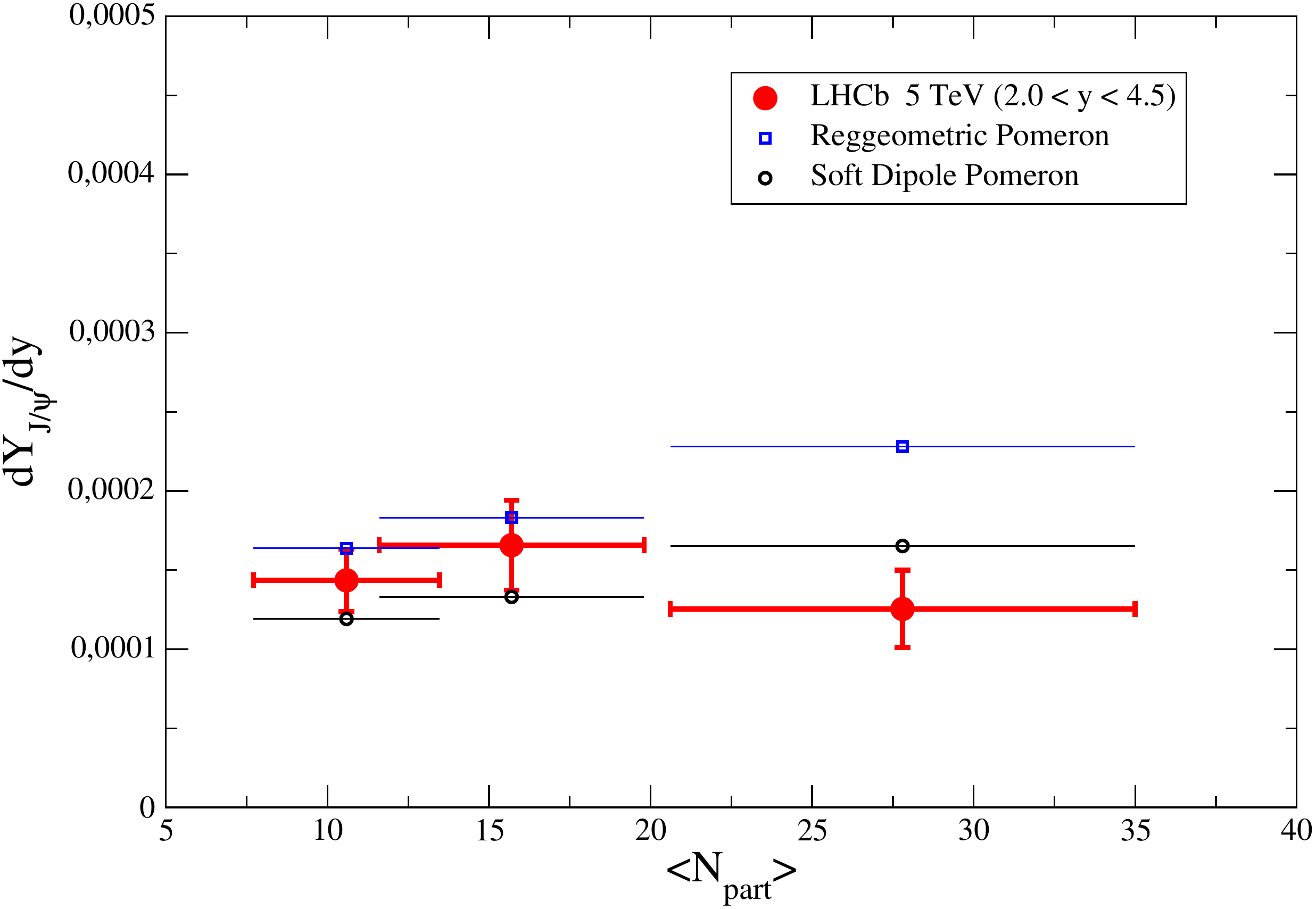}
 \caption{Differential yields of photo-produced $J/\psi$ as a function of number of participants nucleons,  $\langle N_{part} \rangle$, at $\sqrt{s_{\mathrm{NN}}} = 5.02$ TeV in the rapidity range $2.0<y<4.5$. Predictions of  the  Reggeometric Pomeron model (open squares) and Soft Dipole Pomeron model (open circles) are compared to data from LHCb collaboration \cite{LHCb:2021hoq}.}
  \label{fig:NpartLHCb502TeV}
\end{figure}

\begin{figure}[t]
\centering
 \includegraphics[width=.6\textwidth]{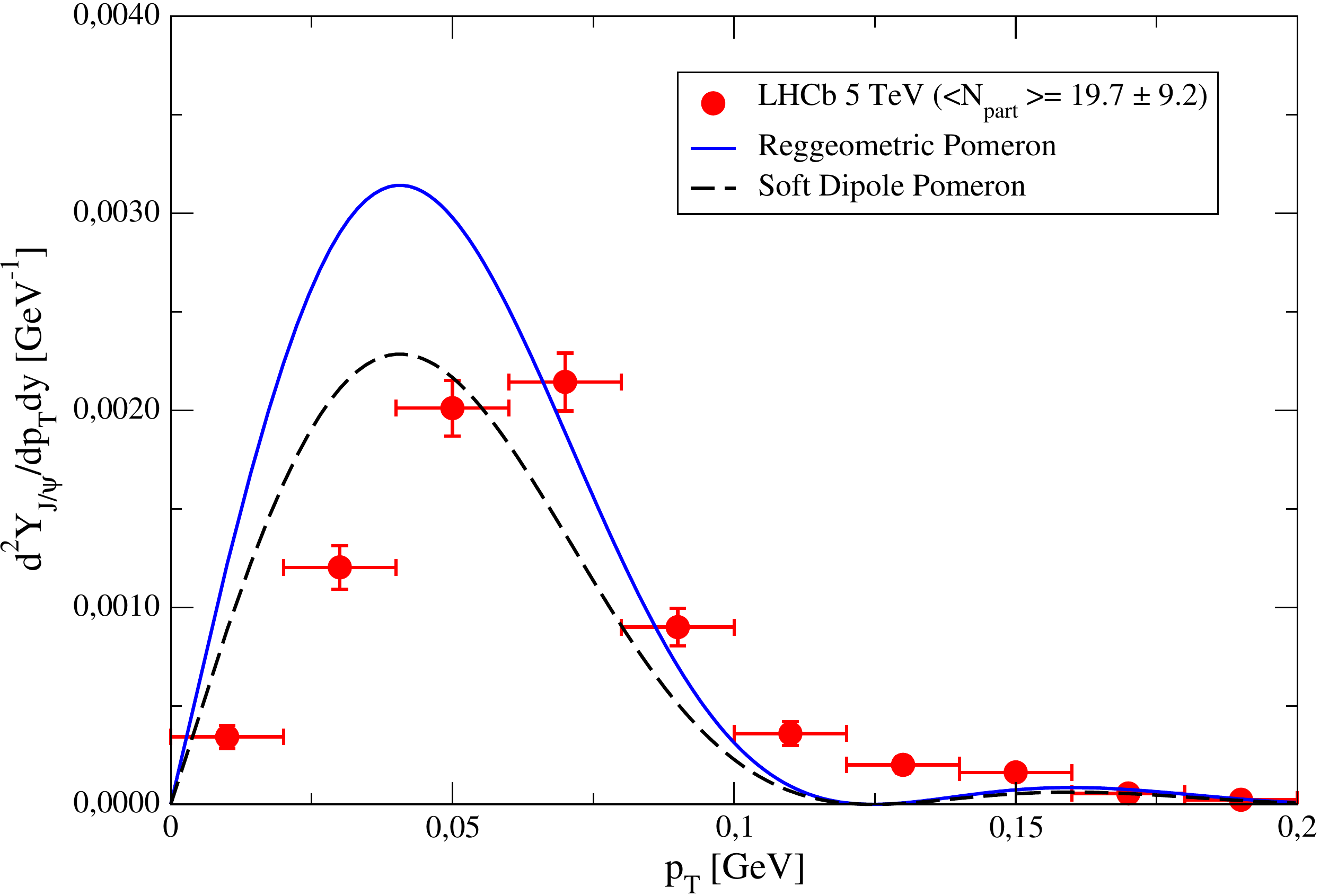}
 \caption{ The double differential $J/\psi$ photoproduction yields as a function of the transverse momentum for $\sqrt{s_{\mathrm{NN}}} = 5.02$ TeV in the rapidity range $2.0<y<4.5$. Predictions of  the  Reggeometric Pomeron model (solid curve) and Soft Dipole Pomeron model (long-dashed curve) are compared to data from LHCb collaboration \cite{LHCb:2021hoq}.}
  \label{fig:dsdpt502TeV}
\end{figure}

Motivated by the recent LHCb measurement for photoproduction of $J/\psi$ at low transverse momentum in peripheral PbPb  collisions at 5 TeV \cite{LHCb:2021hoq}, we calculate the rapidity distribution for fixed  average number of participant.  This is presented in Fig. \ref{fig:LHCB502centrality} with $\langle N_{part} \rangle  = 19.7 \pm 9.2 $ in the rapidity range $ 2.0 < y < 4.5$. In this case both models are consistent with data within the experimental errors. The Soft Dipole Pomeron model seems to be preferred though, similarly to Figures~\ref{fig:ALICE276centrality} and~\ref{fig:ALICE502centrality}. Moreover, in Fig. \ref{fig:NpartLHCb502TeV} is shown the differential yields of photo-produced $J/\psi$ as a function of number of participants nucleons,  $\langle N_{part} \rangle$, in the same rapidity interval. The numerical results for  the  Reggeometric Pomeron model are represented by the open square symbols whereas the ones for  Soft Dipole Pomeron model are labeled by open circle symbols. As expected, the present approach is reasonable for more peripheral collisions (the photon flux has been not modified to include overlap or inteference effects). The predictions are compared to data from LHCb collaboration \cite{LHCb:2021hoq}. The number of binary nucleon-nucleon collisions and the number of participating nucleons are calculated for a given impact parameter using the Glauber approach in relativistic heavy ion physics \cite{Miller:2007ri}.  The input for the nuclear density is given by the Woods- Saxon distribution for lead nuclei. The centrality percentile intervals follows the  supplementary material and additional information  available in the LHCb public pages \cite{LHCB-PAPER-2020-043-webpage}.

As a final investigation, we present the predictions for the transverse momentum distribution for coherent $J/\psi$ photoproduction and compare them to the double differential photoproduction yields measured by LHCb collaboration \cite{LHCb:2021hoq} in PbPb collisions at $\sqrt{s_{\mathrm{NN}}} = 5$ TeV. The $p_T$-distribution is obtained theoretically from the momentum transfer distribution as follows:
\begin{eqnarray}
\frac{d^2\sigma (A+A\rightarrow A+V+A)}{dydp_T^2}=\omega_+\frac{dN_{\gamma/A}(\omega_+)}{d\omega}\frac{d\sigma_{\gamma A\rightarrow VA}(\omega_+)}{d|t|} +\omega_-\frac{dN_{\gamma/A}(\omega_-)}{d\omega}\frac{d\sigma_{\gamma A\rightarrow VA}(\omega_-)}{d|t|},
\end{eqnarray}
where the $t$-dependence is obtained by Eq. (\ref{eq:sigtotgammaA}) after removing the $|t|$-integration. In Fig. \ref{fig:dsdpt502TeV} the results are presented and compared to LHCb data at the rapidity range $2.0<y<4.5$. The Reggeometric Pomeron model (solid curve) and the Soft Pomeron model (long-dashed curve)  overestimate the data at very low-$p_T$ but provide a better agreement for intermediate transverse momenta values. The Reggeometric Pomeron model was already considered in Ref. \cite{Jenkovszky:2021sis}, where it has been compared with the first measurement of ALICE for $t$-dependence in ultra-peripheral PbPb collisions \cite{ALICE:2021tyx}. Based in that analyzes the overestimation at very small $t$ ($p_T$) would be expected. The $t$-dependence is related to the coupling Pomeron-nucleus and the low-$t$ behavior  may be related
to the two-pion threshold in the non-linear Pomeron trajectory,
manifest as a break in proton-proton differential cross section (pion atmosphere
around the nucleon). We quote for instance  Ref. \cite{Jenkovszky:2017efs} for a detailed discussion on this issue. Our results have similar shape but larger normalization than the predictions of Refs. \cite{Zha:2017jch,Zha:2018jin} as pointed out in the LHCb paper \cite{LHCb:2021hoq}.

\section{Conclusions}

In this paper predictions for coherent $J/\psi$ meson photoproduction in  semi-central and peripheral collisions at the LHC are presented and compared with the recent experimental data collected by different collaborations. The theoretical input is based on Regge phenomenology for the production cross section and the effective photon flux for a given centrality class. We considered two models which describe the exclusive vector meson  measurements for proton targets: i) the single-component Reggeometric Pomeron model and ii) the Soft Dipole Pomeron model.  By using the Glauber multiple scattering model and vector meson dominance in order to obtain the photonuclear cross section,  predictions for semi-central and peripheral  PbPb collisions at 2.76 and 5.02 TeV are presented. Concerning the rapidity distributions, the models are in a reasonable agreement with the peripheral collisions especially for forward rapidities.  We provide predictions for the differential yields as a function of number of participants nucleons,transverse momentum and rapidity which are in agreement with experimental results  for peripheral classes of centrality.

 \section*{Acknowledgements}
L.J. was supported by the NASU grant 1230/22-1 \textit{Fundamental Properties of Matter}.
MVTM was supported by funding agencies CAPES (Finance Code 001) and CNPq (grant number 306101/2018-1), Brazil.

\end{document}